\documentclass[preprint
 ,secnumarabic%
,amssymb, amsmath,nobibnotes,prd]{revtex4}
\begin{document}
\title{Noncommutative Topological Theories of Gravity}

\author{H. Garc\'{\i}a-Compe\'an}
\email{compean@fis.cinvestav.mx}
\affiliation{Departamento de F\'{\i}sica,
Centro de Investigaci\'on y de Estudios Avanzados del IPN\\
P.O. Box 14-740, 07000 M\'exico D.F., M\'exico}
\author{O. Obreg\'on}
\email{octavio@ifug3.ugto.mx}
\altaffiliation[Permanent adress:]{
Instituto de F\'{\i}sica de la Universidad de \\
\vskip -1truecm
Guanajuato, P.O. Box E-143, 37150 Le\'on Gto., M\'exico.}
\affiliation{D.A.M.T.P.,
Cambridge University,
Wilberforce Road,
Cambridge CB3 0WA, U.K.}
\author{C. Ram\'{\i}rez}
\email{cramirez@fcfm.buap.mx}
\altaffiliation[Permanent adress:]{
Facultad de Ciencias F\'{\i}sico Matem\'aticas,\\
\vskip -1truecm
Universidad Aut\'onoma de Puebla,
P.O. Box 1364, 72000 Puebla, M\'exico.}
\affiliation{Instituto de F\'{\i}sica de la Universidad de Guanajuato,\\
P.O. Box E-143, 37150 Le\'on Gto., M\'exico}
\author{M. Sabido}
\email{msabido@ifug3.ugto.mx}
\affiliation{Instituto de F\'{\i}sica de la Universidad de Guanajuato\\
P.O. Box E-143, 37150 Le\'on Gto., M\'exico}

\date{\today}

\begin{abstract}

The possibility of noncommutative topological gravity arising in the same
manner as Yang-Mills theory is explored. We use the Seiberg-Witten map to
construct such a theory based on a SL(2,{\bf C}) complex connection, from
which the Euler characteristic and the signature invariant are obtained.
This gives us a way towards the description of noncommutative
gravitational instantons as well as noncommutative local gravitational
anomalies.

\end{abstract}
\vskip -1truecm
\maketitle

\vskip -1.3truecm
\newpage

\setcounter{equation}{0}

\section{Introduction}

The idea of the noncommutative nature of space-time coordinates is quite old \cite{ref1}.
Many authors have extensively studied it from the mathematical \cite{ref2}, as well as
field theoretical points of view (for a review, see for instance \cite{douglas,szabo}).

Recently, noncommutative gauge theory has attracted a lot of attention, specially in
connection with M(atrix) \cite{connes} and string theory \cite {ref3}. In particular,
Seiberg and Witten \cite{ref3} have found noncommutativity in the description of the low
energy excitations of open strings (possibly attached to D-branes) in the presence of a NS
constant background $B-$field. Moreover, they have observed that, depending on the
regularization scheme of the two dimensional correlation functions, Pauli-Villars or point
splitting, ordinary and noncommutative gauge fields can be induced from the same worldsheet
action. Thus, the independence of the regularization scheme tells us that there is a
relation of the resulting theory of noncommutative gauge fields, deformed by the Moyal
star-product, or Kontsevich star product for systems with general covariance, with a gauge
theory in terms of usual commutative fields. This relation is the so-called Seiberg-Witten
map.

In string theory, gravity and gauge theories are realized in very different ways. The
gravitational interaction is associated with a massless mode of closed strings, while
Yang-Mills theories are more naturally described in open strings or in heterotic string
theory. Furthermore, as mentioned, string theory predicts a noncommutative effective
Yang-Mills theory. Thus the question emerges, whether a noncommutative description of
gravity would arise from it. This is a difficult question and it will not be addressed
here. However, in a recent paper \cite{ardalan}, gravitation on noncommutative D-branes has
been discussed.

In this context, recently Chamseddine has made several proposals for noncommutative
formulations of Einstein's gravity \cite{cham,cham1,cham2}, where a Moyal deformation is
done. Moreover, in \cite{cham1,cham2}, he gives a Seiberg-Witten map for the vierbein and
the Lorentz connection, which is obtained starting from the gauge transformations, of
$SO(4,1)$ in the first work, and of $U(2,2)$ in the second one. However, in both cases the
actions are not invariant under the full noncommutative transformations. Namely, in
\cite{cham1} the action does not have a definite noncommutative symmetry, and in
\cite{cham2} the Seiberg-Witten map is obtained for $U(2,2)$, but the action is invariant
under the subgroup $U(1,1) \times U(1,1).$ These actions deformed by the Moyal product,
with a constant noncommutativity parameter, are not diffeomorphism invariant. However, as
pointed out in these works, \cite{cham1,cham2}, they could be made diffeomorphism
invariant, substituting the Moyal $*_M$-product by the Kontsevich $*_K$-product. For other
recent proposals of noncommutative gravity, see
\cite{moffat,chandia,nishino,nair,klemm,dosdim}.

Further, as shown in \cite{wess1,wess2,wess3,wess4,wess5}, starting from the Seiberg-Witten
map, noncommutative gauge theories with matter fields based on any gauge group can be
constructed. In this way, a proposal for the noncommutative standard model based on the
gauge group product $SU(3) \times SU(2) \times U(1)$ has been constructed \cite{wess6}. In
these developments, the key argument is that no additional degrees of freedom have to be
introduced in order to formulate noncommutative gauge theories. That is, although the
explicit symmetry of the noncommutative action corresponds to the enveloping algebra of the
limiting commutative symmetry group, it is also invariant with respect to this commutative
group, fact made manifest by the Seiberg-Witten map.

In this paper, following these results, we present a first step towards a noncommutative
theory of gravity in four dimensions, fully symmetric under the noncommutative symmetry.

We make a proposal for a noncommutative topological quadratic theory of gravity from which 
the corresponding topological invariants of Riemannian manifolds, the Euler characteristic
and the signature, can be obtained. These invariants should classify gravitational
instantons. Further, in this
context of noncommutative gravity, we explore the notion of gravitational instanton. Other
possible global aspects of noncommutative gravity like gravitational anomalies will be
briefly addressed as well.

The paper is organized as follows. In section 2 we quickly review the noncommutative gauge
theories. In section 3 the main features of topological quadratic gravity are introduced,
for the $SO(3,1)$ gauge group, by means of the complex formulation based on the self-dual
topological quadratic gravity. In section 4 we present noncommutative topological gravity,
with explicit results up to order $\theta^{3}$. In section 5, based in the study of the
global properties of the noncommutative version of the Lorentz and diffeomorphism groups,
we explore the possibility of a definition of noncommutative gravitational instantons, as
well as local gravitational anomalies for a theory of gravity. Finally, section 6 contains
our conclusions.

\vskip 1truecm

\section{Noncommutative Gauge Symmetry and the Seiberg-Witten Map}

We start this section with conventions and properties of
noncommutative spaces for future reference. For a recent review see e.g. \cite{zachos}.
Noncommutative spaces can be understood as generalizations of the usual
quantum mechanical commutation relations, by the introduction of a linear operator algebra
${\cal A}$, with a noncommutative associative product,
\begin{equation}
[ \widehat x^{\mu},\widehat x^{\nu} ] =i\theta ^{\mu \nu},
\label{comm}
\end{equation}
where $\widehat x^{\mu}$ are linear operators acting on the Hilbert space $L^2({\bf R}^{n})$ and
$\theta^{\mu \nu}=-\theta ^{\nu \mu}$ are real numbers. The Weyl-Wigner-Moyal correspondence
establishes (under certain conditions) an isomorphic relation between ${\cal A}$ and the algebra
of functions on ${\bf R}^{n}$, the last with an associative and noncommutative $\star$-product, the Moyal product, given by
\begin{equation}
f(x)\star g(x)\equiv \left[ \exp \bigg(\frac{i}{2}\theta^{\mu \nu}{\frac{\partial
}{\partial \varepsilon^{\mu}}}{\frac{\partial }{\partial \eta^{\nu}}} \bigg) %
f(x+\varepsilon )g(x+\eta )\right] _{\varepsilon =\eta =0}\ \ .
\label{moyal}
\end{equation}
In order to avoid causality problems we will take $\theta^{0\nu} = 0.$

Due to the fact that we will be working with nonabelian groups, we must include also matrix
multiplication, so a $\ast$-product will be used as the matrix
multiplication with $\star$-product. Inside integrals,
this product has the property Tr$\int
f_{1}\ast f_{2}\ast f_{3}\ast \cdots \ast f_{n} = {\rm Tr}\int f_{n}\ast f_{1}\ast f_{2}\ast
f_{3}\ast \cdots \ast f_{n-1}$. In particular, the trace of the integral of the product of two functions has the property that
Tr$\int f_{1}\ast f_{2}={\rm Tr}\int f_{1} f_{2}$.

Let us consider a gauge theory with a hermitian connection, invariant under a symmetry Lie group G,
with gauge fields $A_\mu$,
\begin{equation}
\delta_{\lambda }{A}_{\mu} =\partial _{\mu}{\lambda}+i\left[\lambda,{A}_{\mu}\right],
\label{tress}
\end{equation}
where $\lambda=\lambda^iT_i$, and $T_i$ are the
generators of the Lie algebra ${\cal G}$ of the group G, in the adjoint representation.
These transformations are generalized for the noncommutative theory as,
\begin{equation}
\delta_{\lambda }\widehat{A}_{\mu} =\partial _{\mu}\widehat{\Lambda
}+i\left[\widehat\Lambda\stackrel{\ast}{,}\widehat{A}_{\mu}\right],
\label{trafoanc}
\end{equation}
where the noncommutative parameters $\widehat\Lambda$ have some dependence on $\lambda$
and the connection $A$. The commutators $\left[A\stackrel{\ast}{,}B\right]\equiv A\ast B-B\ast A$
have the correct derivative properties when acting on products of noncommutative fields.

Due to noncommutativity, commutators like
$\left[\widehat\Lambda\stackrel{\ast}{,}\widehat{A}_{\mu}\right]$
take values in the enveloping algebra
of ${\cal G}$ in the adjoint representation, ${\cal U}({\cal G},{\rm ad})$. Therefore,
$\widehat\Lambda$ and the
gauge fields $\widehat{A}_\mu$ will also take values in this algebra.
In general,
for some representation $R$, we will denote ${\cal U} ({\cal G},R)$ the corresponding section of the
enveloping
algebra ${\cal U}({\cal G})$.
Let us write for instance $\widehat\Lambda=\widehat\Lambda^I T_I$ and
$\widehat A=\widehat{A}^I T_I$, then,
\begin{equation}
\left[\widehat\Lambda\stackrel{\ast}{,}\widehat{A}_{\mu}\right]=
\left\{\widehat{\Lambda}^{I}\stackrel{\ast}{,}\widehat{A}_{\mu}^{J}\right\} \left[
T_{I},T_{J}\right] +\left[ \widehat\Lambda^{I}\stackrel{\ast}{,}\widehat{A}_{\mu}^{J}\right]
\left\{ T_{I},T_{J}\right\},
\end{equation}
where $\{A\stackrel{\ast}{,}B\}\equiv A\ast B+B\ast A$ is the noncommutative anticommutator.
Thus all the products of the generators $T_I$ will be needed in order to close the algebra
${\cal U}({\cal G},{\rm ad})$.
Its structure can be obtained by successive computation of commutators and anticommutators
starting from the generators of ${\cal G}$, until it closes,
\begin{equation}
\left[ T_{I},T_{J}\right]=i{f_{IJ}}^KT_{K}, \ \ \ \ \left\{ T^{I},T^{J}\right\}
= {d_{IJ}}^KT_{K}.  \nonumber
\end{equation}

The field strength is defined as
$\widehat{F}_{\mu \nu} =\partial _{\mu}\widehat{A}_{\nu}-
\partial _{\nu}\widehat{A}_{\mu}-i [\widehat{A}_{\mu}\stackrel{\ast}{,}\widehat{A}_{\nu}]$,
hence
it takes also values in ${\cal U}({\cal G},{\rm ad})$. From Eq. (\ref{trafoanc}) it turns out
that,
\begin{equation}
\delta_{\lambda }\widehat{F}_{\mu \nu} =i\left( \widehat{\Lambda }%
\ast \widehat{F}_{\mu \nu}-\widehat{F}_{\mu \nu}\ast\widehat{\Lambda
}\right).
\label{trafoefenc}
\end{equation}
We see that these transformation rules can be obtained from the commutative ones, just by replacing
the ordinary product of smooth functions by the Moyal product, with a suitable
product ordering. This allows constructing in simple way invariant quantities.

If the components of the noncommutativity parameter $\theta$ are constant,
then Lorentz invariance is spoiled. In order to recover it
\cite{cham1,cham2,wess3} one should change the Moyal
star product by the Kontsevich star product $*_K$ \cite{kontsevich}. However, as a result
of the diffeomorphism invariance, for an even dimensional (symplectic) spacetime $X$, there exists a
local coordinate system (which coincides with Darboux's coordinate system) in which
$\theta^{\mu\nu}$ is constant. Therefore, without loss of generality, the Kontsevich product can be
reduced to the Moyal one, which will be used from now on.

The fact that the observed world is commutative, means that there must be possible
to obtain it from the noncommutative one by taking the limit $\theta\rightarrow 0$.
Thus the noncommutative fields $\widehat A$ are given by a power series expansion on $\theta$,
starting from the commutative ones $A$,

\begin{equation}
\widehat{A}=A+\theta^{\mu\nu}A^{(1)}_{\mu\nu}+
\theta^{\mu\nu}\theta^{\rho\sigma}A^{(2)}_{\mu\nu\rho\sigma}+\cdots \ \ .
\label{camposnc}
\end{equation}
The coefficients of this expansion are determined by the Seiberg-Witten map, which
states that the symmetry transformations of (\ref{camposnc}), given by (\ref{trafoanc}), are induced
by the symmetry
transformations of the commutative fields (\ref{tress}). In order that these transformations be
consistent,
the transformation parameter $\widehat\Lambda$ must satisfy \cite{wess2},
\begin{equation}
\delta_\lambda\widehat\Lambda(\eta)-\delta_\eta\widehat\Lambda(\lambda)-
i[\widehat\Lambda(\lambda)\stackrel{*}{,}\widehat\Lambda(\eta)]=
\widehat\Lambda(-i[\lambda,\eta]).\label{parametros}
\end{equation}

Similarly, the coefficients in Eq. (\ref{camposnc}) are functions
of the commutative fields and their derivatives, and are determined by the requirement that
$\widehat A$ transforms as (\ref{trafoanc}), \cite{wess5}.

The fact that the noncommutative gauge fields take values in the enveloping algebra, has the
consequence that they have a bigger number of components than the
commutative ones, unless the enveloping algebra coincides with the Lie algebra of the
commutative theory, as is the case of $U(N)$. However, the
physical degrees of freedom of the noncommutative fields can be related one to one to the
physical degrees of freedom of the commutative fields by the Seiberg-Witten map
\cite{ref3}, fact used in references \cite{wess1,wess2,wess3,wess4,wess5} to construct noncommutative
gauge
theories, in principle for any Lie group.

In order to obtain the Seiberg-Witten map to first order, the noncommutative parameters are
first obtained from Eq. (\ref{parametros}) \cite{ref3,wess1,wess2,wess3,wess4,wess5},

\begin{equation}
\widehat{\Lambda }\left( \lambda ,A\right) =\lambda +\frac{1}{4}\theta ^{\mu \nu}\left\{
\partial
_{\mu}\lambda ,A_{\nu}\right\} +{\cal O}\left( \theta ^{2}\right).
\label{difeq}
\end{equation}

Then, from Eqs. (\ref{trafoanc}) and (\ref{camposnc}), the following solution is given

\begin{equation}
\widehat{A}_{\mu}\left( A\right) =A_{\mu}-\frac{1}{4}\theta ^{\rho \sigma}\left\{
A_{\rho},\partial
_{\sigma}A_{\mu}+F_{\sigma \mu}\right\} +{\cal O}\left( \theta ^{2}\right) ,
\label{asw}
\end{equation}
and then for the field strength it turns out that,

\begin{equation}
\widehat{F}_{\mu \nu} =F_{\mu \nu}+\frac{1}{4}\theta ^{\rho \sigma}\bigg( 2\left\{
F_{\mu \rho},F_{\nu \sigma}\right\} -\left\{ A_{\rho},D_{\sigma}F_{\mu \nu}+\partial
_{\sigma}F_{\mu \nu}\right\}
\bigg) +{\cal O}\left( \theta ^{2}\right).
\label{swf}
\end{equation}

The higher coefficients in Eq. (\ref{camposnc}) can be obtained from the observation that the
Seiberg-Witten map preserves the operations of the commutative function algebra,
hence the following differential equation can be written \cite{ref3},
\begin{equation}
\delta\theta^{\mu\nu}\frac{\partial}{\partial\theta^{\mu\nu}}\widehat A(\theta)=
\delta\theta^{\mu\nu}\widehat{A^{(1)}_{\mu\nu}}(\theta),\label{e}
\end{equation}
where $\widehat{A^{(1)}_{\mu\nu}}$ is obtained from ${A^{(1)}_{\mu\nu}}$ in
Eq. (\ref{camposnc}), by substituting the commutative fields by
the noncommutative ones under the $\ast$-product.

Let us take the generators $T^i$ of the Lie algebra ${\cal G}$ to be hermitian, then the generators
$T^I$ of the corresponding enveloping algebra can be chosen to be also hermitian, for instance if
they are given by the symmetrized products $:T^{i_1}T^{i_2}\cdots T^{i_n}:\ $. Further,
the noncommutative transformation parameters $\widehat\Lambda(\lambda,A)$ are functions, whose
arguments are matrices. Let us now substitute the matrix products inside
$\widehat\Lambda(\lambda,A)$, by
$MN\rightarrow \frac{1}{2}\{M,N\}-\frac{i}{2}(i[M,N])$, for any two matrices $M$ and $N$.
Hence $\widehat\Lambda(\lambda,A)$ can be understood as a function
whose nonlinear part of depends polynomially, with complex numerical coefficients,
on anticommutators $\{\cdot,\cdot\}$ and commutators $i[\cdot,\cdot]$, of $\lambda$, $A$,
and their derivatives. With this understanding, we will continue to write it as
$\widehat\Lambda(\lambda,A)$, and we have
\begin{equation}
[\widehat\Lambda(\lambda,A)]^\dagger=
\widehat\Lambda^\dagger(\lambda^\dagger,A^\dagger),
\label{dagger}
\end{equation}
where $\widehat\Lambda^\dagger$ is obtained by complex conjugating the mentioned
numerical coefficients.

Let us now consider the hermitian conjugation of the
transformation law (\ref{tress}),
$(\delta_{\lambda }{A}_{\mu})^\dagger =\partial _{\mu}{\lambda}^\dagger+
i\left[\lambda^\dagger,{A}_{\mu}^\dagger\right]$.
From it and (\ref{parametros}), taking into account (\ref{dagger}),
we get,
\begin{equation}
\delta_{\lambda^\dagger}
\widehat\Lambda^\dagger(\lambda^\dagger,A^\dagger)-
\delta_{\eta^\dagger}\widehat\Lambda^\dagger(\lambda^\dagger,A^\dagger)-
i[\widehat\Lambda^\dagger(\lambda^\dagger,A^\dagger)\stackrel{*}{,}
\widehat\Lambda^\dagger(\eta^\dagger,A^\dagger)]=
\widehat\Lambda^\dagger(-i[\lambda^\dagger,\eta^\dagger],A^\dagger).\label{parametrosnc}
\end{equation}
Comparing this equation with (\ref{parametros}), with the mentioned convention,
it can be seen that the noncommutative parameters satisfy
$[\widehat\Lambda(\lambda,A)]^\dagger=\widehat\Lambda(\lambda^\dagger,A^\dagger)$.
From the transformation law (\ref{trafoanc}), a similar conclusion can be
obtained for the noncommutative connection,
$[\widehat{A}_{\mu}(A)]^\dagger=\widehat{A}_{\mu}(A^\dagger)$,
as well for the field strength,
$[\widehat{F}_{\mu\nu}(A)]^\dagger=
\widehat{F}_{\mu\nu}(A^\dagger)$. By this means,
if we have a group with real parameters and hermitian generators, with a hermitian connection,
then the noncommutative connection
and the noncommutative field strength will be also hermitian.

\section{Topological Gravity}

In this section we shortly review four-dimensional topological gravity. We start from
the following $SO(3,1)$ invariant action
\begin{equation}
I_{TOP}=\frac{\Theta _{G}^{P}}{2\pi }{\rm Tr}\int_{X}R\wedge R+
i\frac{\Theta _{G}^{E}}{2\pi }{\rm Tr}\int_{X}R\wedge \widetilde{R},
\label{topo}
\end{equation}
where $R$ is the field strength, corresponding to a $SO(3,1)$ connection $\omega$
\begin{equation}
R_{\mu \nu }^{\ \ ab}=\partial _{\mu }\omega _{\nu }^{\ ab}-\partial
_{\nu }\omega _{\mu }^{\ ab}+\omega_{\mu }^{\ ac}\omega_{\nu\, c}^{\ \ b}-
\omega_{\mu }^{\ bc}\omega_{\nu\, c}^{\ \ a},\label{riemann}
\end{equation}
$X$ is a four dimensional closed pseudo-Riemannian
manifold and
$\widetilde{ R}_{\mu \nu }^{\ \ \ ab}=-\frac{i}{2}{\varepsilon^{ab}}_{cd}{R_{\mu \nu }}^{cd}$
is the dual with respect to the group.
Here the coefficients are the gravitational analogs of the $\Theta -$vacuum in QCD
\cite{ref7,ref8,ref9}.

In this action, the connection satisfies the first Cartan structure equation, which relates
it to a given tetrad. This action can be written as the integral of a divergence, and a
variation of it with respect to the tetrad vanishes, hence it is metric independent, and
therefore topological.

The action (\ref{topo}) arises naturally from
the MacDowell-Mansouri type action \cite{ref10}. A similar construction can be done
for $(2+1)$-dimensional Chern-Simons gravity \cite{ref11}. Keeping this
philosophy in mind, action (\ref{topo}) can be
rewritten in terms of the self-dual and anti-self-dual parts,
$R^\pm=\frac{1}{2}(R\pm\widetilde R)$, of the Riemann tensor as follows:
\begin{equation}
I_{TOP}={\rm Tr}\int_{X}\left( \tau R^{+}\wedge R^{+}+\overline\tau R^{-}\wedge R^{-}\right)
={\rm Tr}\int_{X}\left( \tau R^{+}\wedge R^{+}+
\overline\tau\overline{R^{+}}\wedge\overline{R^{+}}\right),\label{topopm}
\end{equation}
where $\tau=\left( \frac{1}{2\pi }\right) \left( \Theta _{G}^{E}+i
\Theta _{G}^{P}\right)$, and the bar denotes complex conjugation.
In local coordinates on $X$, this action can be rewritten as
\begin{equation}
I_{TOP}=2 {\rm R}e\ \tau\int_{X}d^{4}x\ \varepsilon ^{\mu \nu \rho \sigma }
{R_{\mu \nu }^{+}}^{ab}R_{\rho \sigma ab}^{+},
\label{selftopo}
\end{equation}
Therefore, it is enough to study the complex action,
\begin{equation}
I=\int_{X}d^{4}x\ \varepsilon ^{\mu \nu \rho \sigma }
{R_{\mu \nu }^{+}}^{ab}R_{\rho \sigma ab}^{+}.
\label{selftopo1}
\end{equation}

Further, the self-dual Riemann tensor satisfies,
${\varepsilon^{ab}}_{cd}{R_{\mu \nu }^{+}}^{cd}= 2i{R_{\mu \nu}^{+}}^{ab} $.
This tensor has the useful property that it can be written as a usual Riemann tensor,
but in terms of the self-dual components of the spin connection,
$\omega _{\mu }^{+ \ ab}=\frac{1}{2}\left( \omega _{\mu }^{ab}- \frac{i}{%
2}{\varepsilon^{ab}}_{cd}\omega _{\mu }^{cd}\right) $, as

\begin{equation}
R_{\mu \nu }^{+ \ ab}=\partial _{\mu }\omega _{\nu }^{+ \ ab}-\partial
_{\nu }\omega _{\mu }^{+ \ ab}+\omega_{\mu }^{+\ ac}\omega_{\nu\ c}^{+\ b}-
\omega_{\mu }^{+\ bc}\omega_{\nu\ c}^{+\ a}.\label{riemannpm}
\end{equation}
In this case, the action (\ref{selftopo}) can be rewritten as,
\begin{equation}
I=\int_{X}d^{4}x\ \varepsilon ^{\mu \nu \rho \sigma }\left[
2{R_{\mu \nu }}^{0i}(\omega^{+})R_{\rho \sigma 0i}(\omega^{+})+
{R_{\mu \nu }}^{ij}(\omega^{+})R_{\rho \sigma ij}(\omega^{+})\right].
\end{equation}
Now, we define ${\omega_\mu}^i=i\omega_{\mu }^{+ 0i}$, from which we obtain,
by means of the self-duality properties,
$\omega_\mu^{+ij}=-{\varepsilon^{ij}}_k{\omega_{\mu}}^k$. Then it turns out that
\begin{eqnarray}
R_{\mu \nu }^{\ \ oi}(\omega ^{+}) &=&-i(\partial _{\mu }\omega _{\nu
}^{i}-\partial _{\nu }\omega _{\mu }^{i}+2 \varepsilon _{jk}^{i}\omega _{\mu
}^{j}\omega _{\nu c}^{k})=-i{\cal R}_{\mu \nu }^{\ \ i}(\omega ) \\
R_{\mu \nu }^{\ \ ij}(\omega ^{+}) &=&\partial_\mu\omega_\nu^{+ij}-\partial_\nu\omega_\mu^{+ij}-
2(\omega_\mu^{\ i}\omega_\nu^{\ j}-\omega_\nu^{\ i}\omega_\mu^{\ j})=
-\varepsilon^{ij}_{\ \ k}{\cal R}_{\mu \nu}^{\ \ k}(\omega ).
\end{eqnarray}
This amounts to the decomposition $SO(3,1)=SL(2,{\bf C}) \times SL(2,{\bf C})$, such that
$\omega_\mu^{\ i}$ is a complex $SL(2,{\bf C})$ connection. If we choose the algebra
$s\ell(2,{\bf C})$ to satisfy $[T_i,T_j]=2i\varepsilon_{ij}^{\ \,k}T_k$ and
Tr$(T_iT_j)=2\delta_{ij}$, then we can write
\begin{equation}
I={\rm Tr}\int_{X}d^{4}x\ \varepsilon^{\mu \nu \rho \sigma}
{\cal R}_{\mu \nu }(\omega){\cal R}_{\rho \sigma}(\omega),
\label{accion2c}
\end{equation}
where, ${{\cal R}_{\mu\nu}}=\partial _{\mu }\omega _{\nu}-\partial _{\nu }\omega _{\mu }-i[\omega _{\mu},\omega _{\nu}]$ is the field strength.
This action is invariant under the $SL(2,{\bf C})$ transformations,
$\delta_\lambda\omega_\mu=\partial_\mu\lambda+i[\lambda,\omega_\mu]$.

In the case of a Riemannian manifold $X$, the signature and the Euler topological invariants of $X$,
are the real and imaginary parts of (\ref{accion2c})
\begin{eqnarray}
\sigma(X)&=&-\frac{1}{24\pi^2}{\rm Re}\,{\rm Tr}\int_{X}d^{4}x\ \varepsilon^{\mu \nu \rho \sigma}
{\cal R}_{\mu \nu }(\omega){\cal R}_{\rho \sigma}(\omega),\label{sigmac}\\
\chi(X)&=&\frac{1}{32\pi^2}{\rm Im}\,{\rm Tr}\int_{X}d^{4}x\ \varepsilon^{\mu \nu \rho \sigma}
{\cal R}_{\mu \nu }(\omega){\cal R}_{\rho \sigma}(\omega).\label{eulerc}
\end{eqnarray}

\vskip 1truecm

\section{Noncommutative Topological Gravity}
We wish to have a noncommutative formulation of the $SO(3,1)$ action (\ref{topo}). Its first term,
can be straightforwardly made noncommutative, in
the same way as for usual Yang-Mills theory,
\begin{equation}
{\rm Tr}\int_{X}\widehat R\wedge \widehat R.
\label{topop}
\end{equation}
If the $SO(3,1)$ generators are chosen to be hermitian, for example in the spin $\frac{1}{2}$
representation given by $\gamma^{\mu\nu}$, then from the discussion at the end of the second
section, it turns out that $\widehat R_{\mu\nu}$ is hermitian and consequently (\ref{topop})
is real.

Instead, for the second term of (\ref{topo}) such an action cannot be written, because it
involves the
Levi-Civita symbol, an invariant Lorentz tensor, but which is not invariant under the full
enveloping algebra. However, as mentioned at the end of the preceding section, this term can be
obtained from Eq. (\ref{accion2c}).

Thus, in general we will
consider as the noncommutative topological action of gravity,
the $SL(2,{\bf C})$ invariant action,
\begin{equation}
\widehat I={\rm Tr}\int_{X}d^{4}x\ \varepsilon^{\mu \nu \rho \sigma}
\widehat{\cal R}_{\mu \nu }\widehat{\cal R}_{\rho \sigma},
\label{accion2cnc}
\end{equation}
where
${\widehat{\cal R}_{\mu\nu}}=\partial _{\mu }\widehat\omega _{\nu}-
\partial _{\nu }\widehat\omega _{\mu }-i
[\widehat\omega _{\mu}\stackrel{\ast}{,}\widehat\omega _{\nu}]$,
is the $SL(2,{\bf C})$ noncommutative field strength.
This action does not depend on the metric of X. Indeed, as well as the commutative one, it
is given by a divergence,
\begin{equation}
\widehat I={\rm Tr}\ \int_{X}d^{4}x\ \varepsilon^{\mu \nu \rho \sigma}
\partial_\mu\left(\widehat{\omega}_{\nu}\ast\partial_\rho \widehat{\omega}_{\sigma}+
\frac{2}{3}\widehat\omega_\nu\ast\widehat\omega_\rho\ast\widehat\omega_\sigma\right).
\label{accion2cnc3}
\end{equation}
Thus, a variation of (\ref{accion2cnc}) with respect to the noncommutative connection, will vanish identically because of the noncommutative Bianchi identities,
\begin{equation}
\delta_{\widehat\omega} \widehat I=8{\rm Tr}\int\varepsilon^{\mu\nu\rho\sigma}\delta\widehat\omega_\mu\ast
\widehat D_\mu \widehat R_{\rho\sigma}\equiv 0,
\end{equation}
where $\widehat D_\mu$ is the noncommutative covariant derivative.

Further, from the first Cartan structure equation, the SO(3,1) connection, and thus its
$SL(2,{\bf C})$ projection $\omega_\mu^{\ i}$, can be written in terms of
the tetrad and the torsion. Furthermore, from the Seiberg-Witten map, the noncommutative
connection can be written as well as $\widehat\omega(e)$. Therefore, a variation of the action
(\ref{accion2cnc}) with respect to the tetrad of the action, can be written as
\begin{equation}
\delta_{e} \widehat I=8{\rm Tr}\int\varepsilon^{\mu\nu\rho\sigma}\delta_e \widehat\omega_\mu(e)\ast
\widehat D_\mu \widehat R_{\rho\sigma}\equiv 0,
\end{equation}
hence it is topological, as the commutative one.

Thus, we see from (\ref{accion2cnc3}) that, in a $\theta -$power expansion of the action,
each one of the resulting terms will be independent of the metrics, as well as they will be given by
a divergence. Thus, unless these terms vanish identically, they will be topological.
Furthermore, the whole noncommutative action, expressed in terms of the commutative fields by the
Seiberg-Witten map, is invariant under the SO(3,1) transformations. Thus, each term of the expansion
will be also invariant. Thus these terms will be topological invariants.

The action (\ref{accion2cnc}) is not real, as well as the limiting commutative action. Hence,
it is not obvious that the signature (\ref{topop}) will be
precisely its real part. In this case we could neither say that $\widehat\chi(X)$ is given
by its imaginary part. In fact we could only say that $\widehat\chi(X)$ could be obtained from
the difference of (\ref{accion2cnc}) and (\ref{topop}). However, the real and the
imaginary parts of (\ref{accion2cnc}) are invariant under SL(2,{\bf C}) and consequently
under SO(3,1), and thus they are the natural candidates for $\widehat\sigma(X)$
and $\widehat\chi(X)$,
as in (\ref{sigmac}) and (\ref{eulerc}).
In order to write down these noncommutative actions as an expansion in $\theta $, we
will take as generators for the algebra of $SL(2,{\bf C})$, the Pauli matrices.
In this case, to second order in $\theta$,
the Seiberg-Witten map for the Lie algebra valued commutative field strength
${\cal R}_{\mu\nu}={{\cal R}_{\mu\nu}}^{i}(\omega)\sigma_{i}$, is given by
\begin{equation}
\widehat{{\cal R}}_{\mu\nu}={\cal R}_{\mu\nu}+
\theta^{\alpha\beta} {\cal R}_{\mu\nu\alpha\beta}^{(1)}+
\theta^{\alpha\beta}\theta^{\gamma\delta}{\cal R}_{\mu\nu\alpha\beta\gamma\delta}^{(2)}+
 \cdots \ ,  \label{riemann}
\end{equation}
where, from Eq. (\ref{swf}) we get,
\begin{equation}
\theta^{\rho\sigma}{\cal R}_{\mu\nu\rho\sigma}^{(1)}=
\frac{1}{2}\theta^{\rho\sigma}\left[2{\cal R}_{\mu\rho}^{\ \ i}{\cal R}_{\nu\sigma i}-
\omega_{\rho}^{\ \ i}\left(\partial_\sigma {\cal R}_{\mu\nu i}+
D_\sigma{\cal R}_{\mu\nu i}\right)\right]{\bf 1} , \label{r1}
\end{equation}

where ${\bf 1}$ is the unity 2$\times$2 matrix. Further, by means of Eq. (\ref{e}), we get,
\begin{eqnarray}
\theta^{\rho\sigma}\theta^{\tau\theta}{\cal R}_{\mu\nu\rho\sigma\tau\theta}^{(2)}=
\frac{1}{4}\theta^{\rho\sigma}\theta^{\tau\theta}\bigg(\varepsilon^i_{jk}\left[i\partial_\tau
{\cal R}^j_{\mu\rho}\partial_\theta {\cal
R}^k_{\nu\sigma}+\partial_\tau\omega^j_\rho \partial_\theta
(\partial_\sigma+D_\sigma){\cal R}^k_{\mu\nu}\right]\nonumber\\
-\omega^i_\rho\partial_\tau\omega^j_\sigma\partial_\theta {\cal R}_{\mu\nu j}+
{\cal R}^i_{\mu\rho}[2{\cal R}^j_{\nu\tau}{\cal R}_{\sigma\theta j}-
\omega^j_\tau(\partial_\theta+D_\theta){\cal R}_{\nu\sigma j}]\nonumber\\
-{\cal R}^i_{\nu\rho}\left[2{\cal R}^j_{\mu\tau}{\cal R}_{\sigma\theta j}-
\omega^j_\tau(\partial_\theta+D_\theta){\cal R}_{\mu\sigma j}\right]+
\frac{1}{2}\omega^j_\tau(\partial_\theta
\omega_{\rho j}+{\cal R}_{\theta\rho j})(\partial_\sigma+D_\sigma){\cal R}^i_{\mu\nu}\nonumber\\
-2\omega^i_\rho
\left\{2\partial_\sigma {\cal R}^j_{\mu\tau}{\cal R}_{\nu\theta j}-\partial_\sigma[\omega^j_\tau(\partial_\theta+
D_\theta){\cal R}_{\mu\nu j}]\right\}\bigg)\sigma_i.
\label{r2}
\end{eqnarray}
Therefore, to second order in $\theta$, the action (\ref{accion2cnc}) will be given by,
\begin{equation}
\widehat I={\rm Tr}\ \int_{X}d^{4}x\ \varepsilon^{\mu \nu \rho \sigma}\left[
{\cal R}_{\mu \nu }{\cal R}_{\rho \sigma}+
2 \theta^{\tau\vartheta}{\cal R}_{\mu\nu}{\cal R}^{(1)}_{\rho \sigma \tau \vartheta}+
\theta^{\tau\theta}\theta^{\vartheta\zeta}\left(2{\cal R}_{\mu\nu}
{\cal R}^{(2)}_{\rho\sigma\tau\theta\vartheta\zeta}+{\cal R}^{(1)}_{\mu\nu\tau\theta}
{\cal R}^{(1)}_{\rho\sigma\vartheta\zeta}\right)\right].
\label{accion2cnc1}
\end{equation}
Taking into account (\ref{r1}), we get that the first order term is proportional to
Tr$(\sigma_i)$ and thus vanishes identically. Further using (\ref{r2}), we finally get,

\begin{eqnarray}
\widehat I&=& \int_{X}d^{4}x\ \varepsilon^{\mu \nu \rho \sigma}\bigg\{
2{\cal R}^i_{\mu \nu }{\cal R}_{\rho \sigma i}+
\frac{1}{4}\theta^{\tau\theta}\theta^{\vartheta\zeta}
\bigg[-\varepsilon_{ijk}R^i_{\mu\nu}\left[
\partial_\vartheta R^j_{\rho\tau}\partial_\zeta R^k_{\sigma\theta}-
\partial_\vartheta\omega^j_\tau\partial_\zeta(\partial_\theta+D_\theta)R^k_{\rho\sigma}\right]
\nonumber\\
&+&[R^i_{\mu\tau}R_{\nu\theta i}-
\frac{1}{2}\omega^i_{\tau}(\partial_\theta+D_\theta)R_{i \mu\nu}]
[R^j_{\rho\vartheta}R_{\sigma\zeta j}-
\frac{1}{2}\omega^j_\vartheta(\partial_\zeta+D_\zeta)R_{\rho\sigma j}]\nonumber\\
&+&R^i_{\mu\nu}\big\{R_{i \sigma\theta}[2R^j_{\rho\vartheta}R_{\tau\zeta
j}-\omega^j_\vartheta
(\partial_\zeta+D_\zeta)R_{\rho\tau j}]+\frac{1}{4}(\partial_\theta+D_\theta)R_{\rho\sigma i}
\omega^j_\vartheta(\partial_\zeta\omega_{\tau j}+R_{\zeta\tau j})\nonumber\\
&+&\omega_{\theta i}[\partial_\tau(R^j_{\rho\vartheta}R_{\sigma\zeta j})-\frac{1}{2}
\partial_\tau\omega^j_\vartheta(\partial_\zeta+D_\zeta)R_{\rho\sigma j}]\big\}-
\frac{1}{2}R^i_{\mu\nu}\omega_{\tau i}\partial_\vartheta\omega^j_\theta\partial_\zeta
R_{\rho\sigma j}\bigg]\bigg\},
\label{accion2cnc2}
\end{eqnarray}
where the second order correction does not identically vanish.

Similarly to the second order term (\ref{r2}), the third order term for $\widehat{\cal
R}$
can be computed by means of Eq. (\ref{e}). The result is given by a rather long expression,
which however is proportional to the unity matrix ${\bf 1}$, like (\ref{r1}).
Thus the third order term in (\ref{accion2cnc1}), given by
\begin{equation}
2\theta^{\tau_1\theta_1}\theta^{\tau_2\theta_2}\theta^{\tau_3\theta_3} {\rm Tr}
\int_{X} \varepsilon^{\mu\nu\rho\sigma}\left({\cal R}_{\mu\nu}
{\cal R}^{(3)}_{\rho\sigma\tau_1\theta_1\tau_2\theta_2\tau_3\theta_3}+
{\cal R}^{(1)}_{\mu\nu\tau_1\theta_1}{\cal R}^{(2)}_{\rho\sigma\tau_2\theta_2\tau_3\theta_3}\right),
\end{equation}
vanishes identically, because ${\cal R}^{(2)}$ is proportional to $\sigma_i$. Thus,
(\ref{accion2cnc2}) is valid to third order. In fact, it seems that all its
odd order terms vanish.

\vskip 1truecm

\section{Towards Noncommutative Gravitational Instantons and Anomalies}

\subsection{Towards Noncommutative Gravitational Instantons}

In the Euclidean signature, the action (\ref{topo}), with local Lorentz group $SO(4)$, is
proportional to a linear combination of integer valued topological invariants, the Euler $\chi(X)$
and the signature $\sigma(X)$, which characterize the gravitational instantons. In fact,
$\sigma(X)$ and $\chi(X)$ are the analogue of
the instanton number $k$ of $SU(2)$-Yang-Mills instantons, which is a manifestation of the
gauge group topology, through $k \in \pi_3(SU(2))$. These topological invariants $\chi$ and
$\sigma$, should of course include the corresponding boundary and $\eta$-invariant terms.
Gravitational instantons
are finite action solutions of the self-dual Einstein equations, which are asymptotically
Euclidean \cite{swh}, or asymptotically locally Euclidean (ALE) \cite{gi}, at infinity (for a
review, see \cite{ginstantons}). Then one would ask about the possibility to get gravitational
instanton solutions in noncommutative gravity. The first natural step would be to analyze the
positive action conjecture \cite{pac}, in the context of noncommutative gravity, although it would requires a more complete  version of noncommutative gravity.
However, it is
possible to give some generic arguments, and we will focus on the description of the global
aspects, by analyzing invariants $\chi$ and $\sigma$ in the noncommutative context. In order to
do that, we concentrate in the spin connection dependence, leaving the explicit metrics
for later analysis.

In the previous section, from explicit computations of the noncommutative corrections (in
the noncommutative parameter $\theta$) of the topological invariants (see Eq.
(\ref{accion2cnc2})), we got that they do not vanish at ${\cal O}(\theta^2)$, hence the classical
topological invariants are clearly modified. Thus, the use of the Seiberg-Witten map for the
Lorentz group leads to essentially modified invariants $\widehat{\chi}$ and $\widehat{\sigma}$,
which would characterize `noncommutative gravitational instantons'. Further, the corresponding deformed equation under the Seiberg-Witten map, $\widehat{R}^+_{\mu \nu}
=0,$ does admit an expansion in $\theta$ with the term at the zero order being $R^+_{\mu \nu}$. Thus these corrections
should be associated to the $\theta -$corrections of the self-duality equation $R^+_{\mu \nu} =
0$. Furthermore, we could expect for the gravitational instantons similar effects
as for the case of Yang-Mills instantons \cite{ns,ref3}, where the singularities of
moduli space are resolved by the noncommutative deformations.

We already know from models of the minisuperspace in quantum cosmology, that noncommutative
gravity leads to a version of noncommutative minisuperspace \cite{ncqc}. Thus, one would expect
some new physical effects from the moduli space of metrics of a noncommutative gravity theory,
which may help to resolve spacetime singularities.

\subsection{Comments on Gravitational Anomalies in Noncommutative Spaces}

\begin{itemize}

\item{\it A Brief Survey on Gravitational Anomalies}

\end{itemize}
The study of topological invariants, leads us also to
other nontrivial topological effects, like the
anomalies, in our gravitational case. Gravitational anomalies,
as well as gauge anomalies, are classified in local and global anomalies. In this paper we will mainly focus on local anomalies,
whereas global anomalies will be mentioned as reference for future work.

Local anomalies are associated to the lack of invariance of the quantum one-loop effective action,
under infinitesimal local transformations.
There are different types of local gravitational anomalies, depending on the type of
transformations, like the Lorentz (or automorphisms) anomaly, and the diffeomorphisms anomaly.

Let ${\cal G}^L_0$ be the group of vertical
automorphisms of the frame bundle over the spacetime $X$. In a local trivialization, the frame
bundle ${\cal
G}^L_0$ can be identified with the set of continuous maps from $X$ to $SO(4)$, which approach to
the identity at infinity, i.e. ${\cal G}^L_0 \equiv Map_0(X,SO(4)) \equiv \{ g: X \to SO(4), \ g
\ {\rm continuous}\}$. Let ${\cal W}$ be the space of gauge field configurations, which consists
of all spin connections $\omega^{ab}_{\mu}(x)$ with appropriate boundary conditions, and let
${\cal B}= {\cal W}/{\cal G}^L_0$. The automorphisms group
${\cal G}^L_0$ acts on ${\cal W}$ in such a way that one can construct the gauge bundle: ${\cal
G}^L_0 \to {\cal W} \buildrel{\pi}\over{\to} {\cal B}$.
In spacetimes $X$ of $n=dim X = 2m$ dimensions, the existence of the local Lorentz gravitational anomaly
is associated to the non-triviality of the non-torsion part of the homotopy of ${\cal B}$, i.e.
$\pi_2({\cal
B}) \cong \pi_1({\cal G}^L_0) = \pi_{2m + 1} (SO(2m))\not= 1.$ For instance for $X = S^4,$ we
get the pure topological torsion $\pi_1({\cal G}^L_0) \cong \pi_5(SO(4)) = \pi_5(SU(2) \times SU(2)) = {\bf Z}_2 \oplus {\bf
Z}_2$. Thus,  in four dimensions there is no local Lorentz anomaly.
However, in $n=4k +2$ dimensions, for $k=0,1,... \ $, it certainly exists.

For local diffeomorphisms transformations,
the moduli space involves a richer phase space structure, given by the quotient space
of the Teichm\"uller space, and the mapping class group.
These anomalies can exist only for $n=4k + 2$ dimensions
for $k=0,1, 2,\cdots$. However, mixed local Lorentz  and diffeomorphism anomalies can exist
in $2k+2$ dimensions \cite{anomaly}.

Global gravitational Lorentz anomalies arise from the fact that  Lorentz
transformations are disconnected, which is related to the nontrivial topology of the group
${\cal G}^L_{\infty} = {\cal G}^L/{\cal G}^L_0$, where ${\cal G}^L$ is the
set of local Lorentz
transformations which have a limit at infinity. In particular for
$X=S^4$, $\pi_0({\cal G}^L_{\infty}) \cong
\pi_4(SO(4)) = \pi_4(SU(2) \times SU(2)) = {\bf Z}_2 \oplus {\bf Z}_2$, and a nontrivial global
Lorentz anomaly arises.
Similarly, the global gravitational diffeomorphisms anomalies are related to the
disconnectedness of the mapping class group $\Gamma^+_{\infty}$, i.e.
$\pi_0(\Gamma^+_{\infty}) \not= 1$ \cite{global}.

\vskip .5truecm

\begin{itemize}

\item{\it Noncommutative Local Lorentz Anomalies}

\end{itemize}

Let us turn to the noncommutative side. The noncommutative version of the the Lorentz group
will be denoted by $\widehat{SO(4)}$, and it is defined in terms of some suitable operator
algebra on a {\it real}\, Hilbert space. Here and in the following, unless otherwise
stated, the noncommutative spaces and groups corresponding to the ones in the preceding
section, will be denoted by hated ones. Following \cite{harvey}, we propose that
$\widehat{SO(4)}$ will be given by the set of compact orthogonal operators ${\bf
O}_{cpt}({\cal H})$, defined on the separable real Hilbert space ${\cal H}$. The
compactness property avoids the Kuiper theorem, which states that the set of pure
orthogonal operators ${\bf O}({\cal H})$ has trivial homotopy groups \cite{kuiper}.
However, the restriction to subalgebras of normed orthogonal operators ${\bf O}_p({\cal H})
= \{ \alpha \ | \ \alpha = {\bf 1} + K\}$ has very important consequences. Here $K$ stands
for compact, finite rank, trace class and Hilbert-Schmidt operator. By a mathematical
result \cite{palais}, the family of normed operator algebras $({\bf O}_{p}({\cal
H}),||\cdot ||_p),$ with the $L^p-$norm given by $||D||_{p} = ({\rm Tr} |D|^p)^{1/p}$,
together with the set $({\bf O}_{cpt}({\cal H}), ||\cdot ||_{\infty}),$ have exactly the
same stable homotopy groups as $SO(\infty)$ (defined through the Bott periodicity theorem).
Further, the stable homotopy groups of $SO(\infty)$, $\pi_j(SO(\infty))$, are given by
${\bf Z}_2$ for $j=0$, ${\bf Z}_2$ for $j=1$, ${\bf Z}$ for $j=3$, and $1$ otherwise. Also
these groups have Bott periodicity mod 8, i.e. $\pi_n(SO(\infty)) = \pi_{n
+8}(SO(\infty))$. Thus, the stable homotopy groups of $\widehat{SO(4)} = {\bf
O}_{cpt}({\cal H})$ are in general nontrivial, and new topological effects in
noncommutative gravity theories are possible.

Le us turn now to the noncommutative analogue of the local Lorentz anomaly.
It is determined by the nontrivial non-torsion
part of homotopy groups of
a suitable  noncommutative version of the Lorentz group $\widehat{\cal G}^L_0$,
which could be defined as the
set $\widehat{\cal G}^L_0 \equiv Map_0(X, {\bf O}_{cpt}({\cal H}))$. The
noncommutative local Lorentz anomaly is detected by the homotopy group
$\pi_2(\widehat{\cal B}) = \pi_1(\widehat{\cal G}^L_0) = \pi_{j}({\bf O}_{cpt}({\cal H}))\not= 1$ for
$j=0,1,3$ mod 8. For $j=0,1$ we have $\pi_{j}({\bf O}_{cpt}({\cal H})) = {\bf Z}_2$, while
for $j=3$, $\pi_{j}({\bf O}_{cpt}({\cal H})) = {\bf Z}$. Thus for $j=3$ a non-torsion
part is detected, and therefore the existence of a local Lorentz anomaly.

\vskip .5truecm
Finally, in the global perspective, the Seiberg-Witten map can be regarded as a map
$SW:{\cal B} \to \widehat{\cal B}$, which preserves the infinitesimal Lorentz transformation
(the gauge equivalence relation), and thus the locally Lorentz invariant observables of the
theory. The
Seiberg-Witten map is not well defined globally since both spaces ${\cal B}$ and
$\widehat{\cal B}$ are different, and their corresponding topologies can be
different as well. However, in some specific cases the operator representation of
the Seiberg-Witten map is quite useful to define the Seiberg-Witten map globally \cite{poly}.

\vskip 1truecm

\section{Concluding Remarks}

In this paper, we propose a noncommutative version for topological gravity with quadratic
actions. We start by the complex action (\ref{accion2cnc}), in terms of the self-dual and
antiself-dual connections, and which contains both the signature and the Euler topological
invariants (for a Riemannian manifold). This action is then written as a $SL(2,{\bf C})$
action, whose noncommutative counterpart can be obtained in the same way as in the
Yang-Mills case, by means of the Seiberg-Witten map. We compute this action up to third
$\theta$-order, and we obtain that the first and the third order vanish, but the second
order is different from zero. The action to this order is given by (\ref{accion2cnc2}). It
seems that all odd $\theta$-orders vanish identically.

The noncommutative signature and the Euler topological invariants are given by the real and
imaginary parts of (\ref{accion2cnc}). For a Riemannian manifold, these topological
invariants characterize gravitational instantons. Thus the study of noncommutative
topological invariants should allow us, through the Seiberg-Witten map, to deform
gravitational instantons into noncommutative versions for them. In order to make explicit
computations, specific gravitational (noncommutative) metrics have to be chosen. In this
context, it would be very interesting to give a noncommutative formulation for dynamical
gravity, following the lines of this work. This analysis will be reported in a forthcoming
paper \cite{ashtekar}.

Similarly to the gauge theories case, we propose a definition of noncommutative
local gravitational Lorentz anomaly, by a suitable definition of the noncommutative
Lorentz group
$\widehat{SO(4)}$ in compact spacetime of Euclidean signature. The application of these
ideas to the diffeomorphism transformations connected to the identity  might predict new
nontrivial noncommutative gravitational
effects, which should be computed explicitly as a noncommutative correction to the
gravitational contribution to the chiral anomaly. The usual gravitational correction was
computed for the standard commutative case in Refs. \cite{salam,anomaly}. Moreover, this
effect can also be regarded as a noncommutative gravitational correction of the local
chiral anomaly in noncommutative gauge theory. This latter case of the pure noncommutative
gauge field was discussed recently in Refs. \cite{ncanomaly}. It would be very interesting
to pursue this way and compare with the results given recently by Perrot \cite{perrot}.

Regarding noncommutative global Lorentz anomalies, in order to understand them, we would
need to specify the connected components of the corresponding group $\widehat{\cal
G}^L_{\infty}$. In this case one would have to compute $\pi_1(\widehat{\cal
W}/\widehat{\cal G}^L_{\infty}) = \pi_0(\widehat{\cal G}^L_{\infty}) \not=1$. Of course a
suitable operator definition of $\widehat{\cal G}^L_{\infty}$ is necessary like in the case
of the local Lorentz anomaly. This is a difficult open problem.

Finally, the ALE gravitational instantons is an important case of gravitational instantons,
which can be obtained as smooth resolutions of {\bf A-D-E} orbifold singularities ${\bf
C}^2 /\Gamma$, with $\Gamma$ being an {\bf A-D-E} finite subgroup of $SU(2)$. These
gravitational instantons are classified through the Kronheimer construction
\cite{kronheimer}, which is the analogue construction to the ADHM construction of
Yang-Mills instantons. There is a proposal to extend the ADHM construction to the
noncommutative case \cite{ns}.  Thus, it would be interesting to give the noncommutative
analogue of the Kronheimer construction of ALE instantons.

\vskip 1truecm
\centerline{\bf Acknowledgments}

This work was supported in part by CONACyT M\'exico Grants Nos. 37851E and
33951E, as well as by the sabbatical grants 020291 (C.R.) and 020331 (O.O.).


\vskip 2truecm




\begin{references}

\bibitem{ref1}  H. Snyder, {Phys. Rev.} {\bf 71} (1947) 38.

\bibitem{ref2}  A. Connes, {\it Noncommutative Geometry}, Academic Press
(1994).

\bibitem{douglas}  M.R. Douglas and N.A. Nekrasov, Rev. Mod. Phys. {\bf 73}
(2002), 977.

\bibitem{szabo}  R.J. Szabo, ``Quantum Field Theory on Noncommutative Spaces'', hep-th/0109162.

\bibitem{connes}  A. Connes, M. R. Douglas, and A. Schwarz, {JHEP} {\bf %
9802:003} (1998).

\bibitem{ref3}  N. Seiberg and E. Witten, {JHEP} {\bf 9909:032} (1999).

\bibitem{ardalan} F. Ardalan, H. Arafaei, M.R. Garousi and Ghodsi, "Gravity on noncommutative D-branes, hep-th/0204117.

\bibitem{cham}  A.H. Chamseddine, {Commun. Math. Phys.} {\bf 218}
(2001) 283.

\bibitem{cham1}  A.H. Chamseddine, {Phys. Lett.} B {\bf 504} (2001) 33.

\bibitem{cham2}  A.H. Chamseddine, ``Invariant Actions for Noncommutative
Gravity'', hep-th/0202137.

\bibitem{moffat}  J.W. Moffat, {Phys. Lett.} B {\bf 491} (2000) 345; {Phys.
Lett.} B {\bf 493} (2000) 142.

\bibitem{chandia}  M. Ba\~{n}ados, O. Chandia, N. Grandi, F.A. Schaposnik
and G.A. Silva, Phys. Rev. D
{\bf 64} (2001) 084012.

\bibitem{nishino}  H. Nishino and S. Rajpoot, Phys. Lett. B {\bf 532} (2002)
334.

\bibitem{nair}  V.P. Nair, ``Gravitational Fields on a Noncommutative
Space'', hep-th/0112114.

\bibitem{klemm}  S. Cacciatori, D. Klemm, L. Martucci and D. Zanon, Phys. Lett. B {\bf 536} (2002) 101.

\bibitem{dosdim}  S. Cacciatori, A.H. Chamseddine, D. Klemm, L. Martucci,
W.A. Sabra and D. Zanon, ``Noncommutative Gravity in Two Dimensions'',
hep-th/0203038.

\bibitem{wess1} J. Madore, S. Schraml, P. Schupp and J. Wess, {Eur. Phys. J. C} {\bf 16} (2000)
161.
\bibitem{wess2} B. Jurco, S. Schraml, P. Schupp and J. Wess, {Eur. Phys. J. C} {\bf 17}
(2000) 521.
\bibitem{wess3} B. Jurco, P. Schupp and J. Wess, {Nucl. Phys.} B {\bf 604}
(2001) 148.
\bibitem{wess4}  J. Wess, {Commun. Math. Phys.} {\bf 219} (2001) 247.

\bibitem{wess5} B. Jurco, L. Moller, S. Schraml, P. Schupp and J. Wess, {Eur.
Phys. J. C} {\bf 21} (2001) 383.

\bibitem{wess6}  X. Calmet, B. Jurco, P. Schupp, J. Wess and M. Wohlgenannt,
{\it Eur. Phys. J. C} {\bf 23} (2002) 363.

\bibitem{zachos} C.K. Zachos, Int. J. Mod. Phys. A {\bf 17} (2002) 297.

\bibitem{kontsevich}  M. Kontsevich, ``Deformation Quantization of Poisson
Manifolds I'', q-alg/9709040.

\bibitem{ref7}  S. Deser, M. J. Duff, and C. J. Isham, {Phys. Lett. B} {\bf %
93} (1980) 419.

\bibitem{ref8}  A. Ashtekar, A. P. Balachandran, and So Jo, {Int. J. Mod.
Phys. A} {\bf 4} (1989) 1493.

\bibitem{ref9}  L. Smolin, {J. Math. Phys.} {\bf 36} (1995) 6417.

\bibitem{ref10}  J. A. Nieto, O. Obreg\'{o}n, and J. Socorro, {Phys. Rev. D}
{\bf 50} (1994) R3583.

\bibitem{ref11}  H. Garc\'{\i}a-Compe\'an, O. Obreg\'{o}n, C. Ram\'{\i}rez, and
M. Sabido, {Phys. Rev. D} {\bf 61} (2000) 085022.

\bibitem{swh} S.W. Hawking, Phys. Lett. A {\bf 60} (1977) 81.

\bibitem{gi} T. Eguchi and A.J. Hanson, Phys. Lett. B {\bf 74} (1978) 249; G.W. Gibbons and
S.W. Hawking, Phys. Lett. B {\bf 78} (1978) 430.

\bibitem{ginstantons} T. Eguchi and A.J. Hanson, Ann. Phys. {\bf 120} (1979) 82; T. Eguchi, P.B.
Gilkey and A.J. Hanson, Phys. Rep. {\bf 66} (1980) 213; {\it Euclidean Quantum Gravity}, eds.
G.W. Gibbons and S.W. Hawking, World Scientific, Singapore (1993).

\bibitem{pac} G.W. Gibbons, S.W. Hawking and M.J. Perry, Nucl. Phys. B {\bf 138} (1978) 141;
G.W. Gibbons and C.N. Pope, Commun. Math. Phys. {\bf 66} (1979) 267; R. Schoen and S.T. Yau,
Phys. Rev. Lett. {\bf 42} (1979) 547; E. Witten, Commun. Math. Phys. {\bf 80} (1981) 381.

\bibitem{ns} N. Nekrasov and A. Schwarz, Commun. Math. Phys. {\bf 198} (1998) 689.

\bibitem{ncqc} H. Garc\'{\i}a-Compe\'an, O. Obreg\'{o}n and C. Ram\'{\i}rez, Phys. Rev. Lett.
{\bf 88} (2002) 161301.

\bibitem{anomaly}  L. Alvarez-Gaum\'e and E. Witten, Nucl.
Phys. B {\bf 234} (1983) 269.

\bibitem{global} E. Witten, Commun. Math. Phys. {\bf 100} (1985) 197.

\bibitem{harvey} J.A. Harvey, ``Topology of the Gauge Group in Noncommutative Gauge Theory'',
hep-th/0105242.

\bibitem{kuiper} N.H. Kuiper, Topology {\bf 3} (1965) 19.

\bibitem{palais} R.S. Palais, Topology {\bf 3} (1965) 271.

\bibitem{poly} P. Kraus and M. Shigemori, ``Noncommutative Instantons and the
Seiberg-Witten Map'', hep-th/0110035; A.P. Polychronakos, ``Seiberg-Witten Map and
Topology'', hep-th/0206013.

\bibitem{ashtekar}  H. Garc\'{\i}a-Compe\'an, O. Obreg\'{o}n, C. Ram\'{\i}rez, and
M. Sabido, to appear.

\bibitem{salam} R. Delbourgo and A. Salam, Phys. Lett. B {\bf 40} (1972)381; T. Eguchi and
P.G.O. Freund, Phys. Rev. Lett. {\bf 37} (1976) 1251.

\bibitem{ncanomaly} F. Ardalan and N. Sadooghi, Int. J. Mod. Phys. A {\bf 16} (2001) 3151;
J.M. Gracia-Bondi and C.P. Martin, Phys. Lett. B {\bf 479} (2000) 321.

\bibitem{perrot} D. Perrot, J. Geom. Phys. {\bf 39} (2001) 82.

\bibitem{kronheimer} P.B. Kronheimer, J. Diff. Geom. {\bf 29} (1989) 665.




\end{references}
\end{document}